\documentclass[journal]{IEEEtran}

\usepackage{graphicx}
\usepackage{amsmath}
\usepackage{algorithm}
\usepackage{subcaption}
\usepackage{color, colortbl}
\definecolor{Gray}{gray}{0.9}

\usepackage[noend]{algpseudocode}

\begin{document}

\title{On-Demand Deployment of Multiple Aerial Base Stations for Traffic Offloading and Network Recovery}

\author{\IEEEauthorblockN{Sanaa Sharafeddine and Rania Islambouli} \\
	\IEEEauthorblockA{\textit{Department of Computer Science and Mathematics}\\
		\textit{Lebanese American University (LAU)}\\
		Beirut, Lebanon\\
		sanaa.sharafeddine@lau.edu.lb}
}
\maketitle

\begin{abstract}
Unmanned aerial vehicles (UAVs) are being utilized for a wide spectrum of applications in wireless networks leading to attractive business opportunities. In the case of abrupt disruption to existing cellular network operation or infrastructure, e.g., due to an unexpected surge in user demand or a natural disaster, UAVs can be deployed to provide instant recovery via temporary wireless coverage in designated areas. A major challenge is to determine efficiently how many UAVs are needed and where to position them in a relatively large 3D search space. 
To this end, we formulate the problem of 3D deployment of a fleet of UAVs as a mixed integer linear program, and present a greedy approach that mimics the optimal behavior assuming a grid composed of a finite set of possible UAV locations. 
In addition, we propose and evaluate a novel low complexity algorithm for multiple UAV deployment in a continuous 3D space, based on an unsupervised learning technique that relies on the notion of electrostatics with repulsion and attraction forces. We present  performance results for the proposed algorithm as a function of various system parameters and demonstrate its effectiveness compared to the close-to-optimal greedy approach and its superiority compared to recent related work from the literature.   
\end{abstract}


\section{Introduction}
Recent research efforts are unlocking the huge potential of unmanned aerial vehicles~(UAVs) in different disciplines and industries including telecommunications, transportation, and security, among others. The value of UAV-based solutions to the telecommunications industry alone is estimated to be around 6.3~billion~USD~\cite{PwC}. UAV-mounted base stations serve as an attractive alternative to provide temporary wireless access to mobile users or sensor devices located in hard-to-reach areas, disaster regions with destroyed network infrastructure, or dense networks with intermittent surge in traffic demand. Recent United Nations and Red Cross reports consider UAVs among the most promising technologies during emergency response and disaster management~\cite{disMngmnt},~\cite{RedCross}, due to advances in their capabilities that include real-time imaging, wireless connectivity, rapid and flexible deployment, high-end computing, and remote sensing. The use of UAVs complements traditional operations in disaster relief by providing fast response especially to areas that impose high risk to victims and first responders~\cite{XZYL2016, Mur2016, SGBW2016, WCLY2017}, and complements other solution approaches for dealing with node failures in wireless cellular and sensor networks~\cite{KK10, YS14, SJ17}.   

In this work, we consider the utilization of UAVs to augment the infrastructure of an existing wireless cellular system upon the occurrence of any event that leads to notable increase in connection outage, for example, concerts or exhibitions that attract a large number of participants in a given geographical location for a limited time period or natural disasters where one or more base stations become out of operation in a given affected area. 

This, however, comes with major technical challenges that include, among others, determining the number of needed UAV-mounted base stations and their locations in a 3D search space while taking into account practical performance aspects such as energy consumption, cost and complexity.  
To this end, we formulate the problem of 3D deployment of multiple UAVs as a mixed integer linear program assuming a grid composed of a finite set of possible locations over several vertical layers. We solve the optimization problem for special case scenarios due to its high complexity and present a corresponding greedy approach that is shown to achieve close-to-optimal performance. We then propose and evaluate a novel low complexity algorithm that allows a fleet of UAVs to be efficiently deployed in a given 3D space to provide connectivity to users who are not covered via the existing wireless network infrastructure. 

The novelty of this work is two-fold: i.~the efficient deployment of a fleet of UAVs in a continuous 3D space whereas the related literature has focused mostly either on single UAV deployment in 3D space or multiple UAV deployment in 2D space (see Section~\ref{sec:related} for details); ii.~the solution approach of the proposed low complexity algorithm that is based on an analogy with electrostatic systems and field forces (see Section~\ref{sec:force} for details). This approach is inspired by research advances on devising supervised and unsupervised learning frameworks based on concepts from dynamic physical fields. The innovation lies in exploring analogies between data points and charged particles, subject to constraints and rules, with applications ranging from classification to clustering problems~\cite{RG09, BG11, SH13}; for example, the authors in~\cite{khandani2009novel} have utilized electrostatic field concepts in order to determine the centroids of data clusters with results demonstrating less sensitivity to noise compared to the standard k-means algorithm, and the authors in~\cite{HC09} have applied electromagnetic field theory concepts to the standard problem of clustering with user constraints. More recently, the authors in~\cite{UN18} have used electrical field dynamics to design optimal generative adversarial networks, which aim at generating synthetic data that cannot be discriminated from an existing training data set. In our work, we model users as fixed charged data points distributed over a given geographical area, and we model UAVs as oppositely charged data points that move dynamically until reaching a steady state in a continuous 3D space based on an electrostatic field composed of attraction forces between UAVs and users in addition to repulsion forces among UAVs.        

In Section~\ref{sec:related}, we cover the related literature and highlight the main contributions of this work. Section~\ref{sec:model} provides the key components of the system model. Section~\ref{sec:formulation} presents an optimization formulation of the problem in addition to a close-to-optimal greedy solution approach with supporting performance results. Section~\ref{sec:force} describes the proposed low complexity  algorithm for traffic offloading using on-demand 3D deployment of multiple UAVs. Section~\ref{sec:results} presents performance results for various scenarios and highlights the effectiveness of the proposed algorithm. Finally, conclusions are drawn in Section~\ref{sec:conc}. 

\section{Related Literature}
\label{sec:related}


Research interest in deploying UAVs as aerial base stations has been growing at a fast pace recently. In~\cite{7412759}, the authors consider two scenarios for using one UAV to cover a given area with uniform distribution of users in the presence of device-to-device communications. The first scenario assumes a static UAV placed at the center of the area of interest and the second assumes a mobile UAV that stops at designated points, called stop points, to provide coverage for the whole area. The authors model the problem as a disc covering problem to determine the number of discs that can fit within the target area, where stop points are selected as the centers of the resulting discs. In both scenarios, they calculate coverage probabilities and achievable system sum bit rates. The authors in~\cite{6564778} investigate the problem of assigning given UAVs as relay nodes to a set of users in wireless ad hoc networks and reformulate the problem as a quadratic unconstrained binary optimization problem. The solution associates each user to one UAV by minimizing the sum of distances between users and their assigned UAVs. 

The authors in~\cite{7762053} propose a polynomial time algorithm to sequentially place UAV-mounted base stations in a spiral manner starting from the outer perimeter of the given area towards the center until all users are covered. Similarly, the authors in~\cite{7461487} demonstrate that UAV-mounted base stations outperform their static picocell counterparts as they can be positioned in a way to maintain direct line of sight with the served users; in order to position the UAVs in locations that complement an existing macrocell infrastructure, the authors propose an algorithm based on k-means clustering.  In~\cite{zhang2017optimal}, the authors address problems related to providing wireless coverage in emergency situations with focus on minimizing metrics related to UAV deployment delays. The authors first propose a low complexity algorithm in the case when all UAVs are dispatched from the same location; then, they extend their work to the more general case of having varying locations by reformulating the min-sum problem as a dynamic program and by proposing a polynomial time approximation scheme to solve it. The authors in~\cite{8046766} consider the problem of deploying a large mesh of UAVs to provide coverage to mobile users distributed in non-uniform densities; they propose a mechanism that starts with an initial deployment of UAVs forming equilateral triangles and then enables UAVs to relocate based on changes in user demand while minimizing the energy consumed in traveling to new positions. In~\cite{7972194}, the authors address the problem of deploying one UAV as a communications infrastructure between two static nodes; they determine the optimal position of the UAV based on the geographical location of the two nodes and their received signal strengths. Some of the papers surveyed in this paragraph address the problem of multiple UAV deployment, however, they assume a 2D search space with common height level for all. 

The authors in~\cite{7510820} study the 3D placement of a single UAV to offload the maximum possible number of users from an existing base station. Further extensions on that work included deploying a single UAV to maximize the coverage of a set of users with different quality of service~(QoS) requirements~\cite{alzenad20173d}. In~\cite{ghanavi2018efficient}, the authors utilized a Q-learning technique to find a 3D position of one aerial base station in order to maximize the aggregate throughput of a terrestrial network while accounting for user mobility. In addition, the authors in~\cite{7417609} derive the optimal altitude of a single UAV to achieve maximum downlink ground coverage and minimum transmit power consumption; this latter work then investigates the optimal deployment of two UAVs to maximize coverage and determine the altitude of the UAVs as well as the distance separating them.   

In light of the surveyed literature, the vast majority of the existing work considers UAV-mounted base stations to be positioned in a 2D plane, with more recent work considering positioning of one or two UAVs in a 3D space. In terms of problem definition, the key novelty of our work is the focus on {\it multiple} UAV deployment over a 3D continuous space taking into account performance and complexity aspects. In terms of solution approach, the key novelty is the proposed efficient low complexity algorithm based on the notion of electrostatic systems and field forces. In terms of application scenarios, the obtained results and insights apply to a wide range of scenarios where unplanned surge in traffic demand takes place due to either increase in user density (e.g., major gathering event) or failure of existing wireless network infrastructure (e.g., disaster event). 

\section{System Model}\label{sec:model}

\begin{figure}[t!]
\centering
\includegraphics[width=0.5\textwidth]{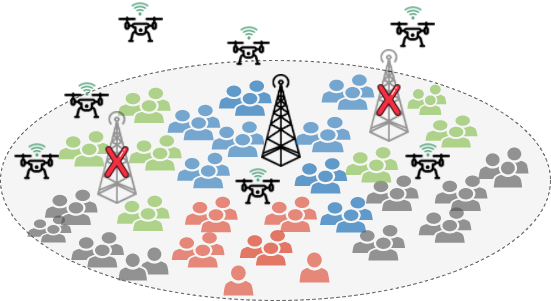}
\caption{Example dense wireless network including one active TBS in addition to a set of ABSs dispatched to fill the coverage gaps resulting from the failure of two TBSs. }
\label{fig:model}
\end{figure}

We consider a fleet of UAV-mounted base stations, denoted as aerial base stations (ABSs), that are dispatched by a network operator as soon as user demand cannot be handled by an existing cellular network infrastructure. We aim at identifying the minimum number of ABSs needed and their optimized 3D locations to provide efficient connectivity to a set of uncovered users. Fig.~\ref{fig:model} presents a sample system model that includes the existence of one terrestrial base station~(TBS) in addition to a number of ABSs to cover a dense area of users. 

\par

We denote by $K_T$ the total number of users in the area and by $N$ the number of needed ABSs. Without loss of generality, we consider the existence of one TBS located in the center of the area, in order to demonstrate how UAVs can dynamically complement an existing cellular infrastructure in dense network scenarios. Due to limited resources, we introduce a capacity constraint $K_B$ on the maximum number of users that can be served by a TBS and a capacity constraint $K_D$ on the maximum number of users that can be served by a single ABS. Based on proper resource allocation, we assume the ABSs and the TBS operate at different frequency bands. 



We consider the following path loss model for the TBS:
\begin{equation}
\mathrm{PL}_{B} =\frac{P_T}{P_R} = \kappa \left(\frac{d}{d_0}\right)^a,
\label{pathloss}
\end{equation}
where $P_R$ is the received power, $P_T$ is the transmitted power, $\kappa$ is the path loss constant, $a$ is the path loss exponent, $d_0$ is a reference distance, and $d$ is the distance in meters between a user and the TBS. As for the ABSs, we adopt the path loss model suggested in \cite{6863654} which depends on the height and angle formed with respect to the served user, resulting in the following line of sight~(LoS) probability: 
\begin{equation}
p_{\mathrm{LoS}}= \frac{1}{1+\mu \cdot \exp\left(-\gamma\left(\arctan\left(\frac{h}{r}\right)-\mu\right)\right)}
\label{LoS}
\end{equation}
\noindent where $h$ and $r$ are the height and the horizontal distance between the ABS and the user, respectively. In addition, $\mu$  and $\gamma$ are constants that depend on the environment. The path loss between an ABS and a user can then be modeled as follows:
\begin{equation}
\begin{aligned}
\mathrm{PL}_{D}  ={} & \frac{P_T}{P_R} = 20\log\left(\frac{4\pi f_c}{c}\right) + 20\log\left(\sqrt{h^2+r^2}\right)\\
      & + p_{\mathrm{LoS}}\eta_{\mathrm{LoS}} + \left(1 - p_{\mathrm{LoS}}\right)\eta_{\mathrm{nLoS}}, 
\end{aligned}
\label{p2}
\end{equation}
\noindent where $f_c$ is the carrier frequency in Hz, $c$ is the speed of light in m/s, $\eta_{LoS}$ and $\eta_{nLoS}$ are, respectively, the losses corresponding to LoS and non-LoS connections depending on the environment. 

\section{UAV Deployment Problem Formulation with Discrete Search Space}
\label{sec:formulation}


In this section, we formulate the problem of UAV deployment for traffic offloading and network recovery assuming a discrete search space with finite set  of $N_D$ possible locations. 
The problem aims at two objectives: 1)~minimize the required number of ABSs in order to reduce cost to the network operator and 2)~deploy the ABSs in positions that maximize the received power for the served users in order to increase the download bit rate and enhance the overall network performance. Table~\ref{params-variables-table} summarizes the system model parameters and decision variables that are used in the problem formulation.

\begin{table}[]
\centering
\caption{System model parameters and variables}
\label{params-variables-table}
\begin{tabular}{|p{1.4cm} |p{6.6cm}|}
\hline
\rowcolor{Gray}
Parameters              & Description                            \\ \hline
$K_T$                   & Total number of users \\ \hline
$K_B$                   & Maximum number of users served by one TBS \\ \hline
$K_D$                   & Maximum number of users served by one ABS \\ \hline
$D$                     & Set of possible sites (ABS and TBS) \\ \hline 
$N_D$                   & Maximum number of possible ABSs \\ \hline
$P_{k,i}$               & Power received at user $k$ from site $i$ \\ \hline
$\Gamma$                & Required SNR target \\ \hline
$\sigma^2$              & Thermal noise power \\ \hline
\rowcolor{Gray}
Variables               & Description \\ \hline
$N$                     & Number of needed ABSs  \\\hline
$\mathcal{A}$           & $K_T\times (N_D+1)$ association matrix between users and sites (ABS or TBS), $a_{k,i}$ is 1 when user $k$ is associated to site $i$ and is 0 otherwise. TBS is denoted as site 0. \\\hline
$b_i$                   & Binary variable that is set to 1 when an ABS is deployed at site $i$ and 0 otherwise \\ \hline
\end{tabular}
\end{table}

\par
\begin{figure}[t!]
\centering
\includegraphics[width=0.45\textwidth]{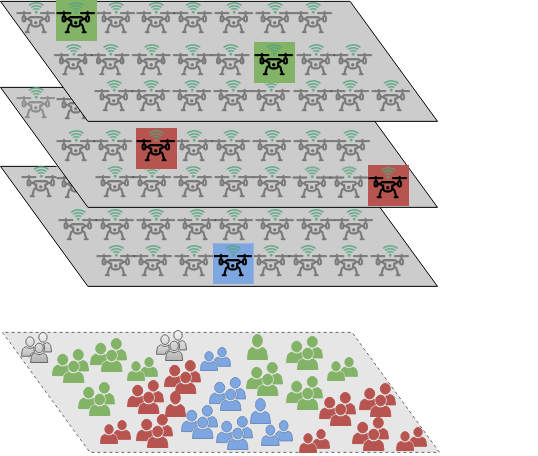}
\caption{Possible locations of ABSs are represented as an array of 2D grids at multiple heights. As an example, selected positions to deploy UAVs are highlighted in bolder colors and the users are colored depending on the color of their serving UAV. Users colored in gray represent those out of coverage.}
\label{fig:UAV_grids}
\end{figure}

To formulate the problem over a 3D search space, the possible positions of ABSs are represented by multiple 2D grids placed at different heights and span the whole area of interest as depicted in Fig.~\ref{fig:UAV_grids}.  
We introduce a binary decision variable $b_i$, for all $1\le i \le N_D$ to indicate whether an ABS is positioned at site $i$ on the grid, as follows: 
\begin{equation}
    b_i=
    \begin{cases}
      1, & \text{if ABS is deployed at site i}\  \\
      0, & \text{otherwise}
    \end{cases}
\end{equation}

 Without loss of generality, we assume the presence of one TBS located in the center at site $0$ and, thus, $b_0 = 1$.  We also define $K_T\times(N_D+1)$ association matrix $\mathbf{A}$ where $a_{ki}$ is a binary variable indicating whether user $k$ is associated with the ABS at position $i$ if $1\le i \le N_D$  or with the TBS if $i=0$.
   
A user is considered in the coverage region of TBS and/or ABS if the received power at the user satisfies a minimum signal to noise ratio~(SNR) constraint. For a given transmission power of TBS/ABS, let $ \Gamma  $ represent the required SNR target to enable user $k$ to associate with site $i$. Thus, 
\begin{equation}
    {P_{k,i}}\geq {\sigma ^2} \Gamma,
\end{equation}
\noindent determines the received power constraint at user $k$ so it can associate with site $i$, where $P_{k,i}$ is the received power level at user $k$ from site $i$ and $\sigma^2 $ is the thermal noise power. 


Given the above, the problem can be formulated as a mixed integer linear program as follows:

\begin{IEEEeqnarray}{l}
	\label{obj_function}
	\displaystyle{\min_{\mathbf{A}}}   \displaystyle\sum\limits_{i=1}^{N_D} { \displaystyle\sum\limits_{k=1}^{K_T}{(\lambda b_i - P_{k,i} a_{k,i})}} \\
	\textrm{Subject to: } \nonumber \\
	\label{constraint_1}
	a_{k,i}\leq b_i ~~\substack{\forall i \in [1,N_D], \\
	\forall k \in [1, K_T]} \\
	\label{constraint_2}
	\displaystyle\sum\limits_{i=1}^{N_D} {a_{k,i}+a_{k,0}} \leq 1 ~~\forall k \in [1, K_T]\\
	\label{constraint_3}
	\displaystyle\sum\limits_{k=1}^{K_T} {a_{k,i}} \leq K_D ~~  \forall i \in [1,N_D]\\
	\label{constraint_4}
	\displaystyle\sum\limits_{k=1}^{K_T} {a_{k,0}} \leq K_B  \\
	\label{constraint_5}
	\displaystyle\sum\limits_{k=1}^{K_T} {\displaystyle\sum\limits_{i=1}^{N_D} {a_{k,i}}}+\displaystyle\sum\limits_{k=1}^{K_T} {a_{k,0}} \geq {(1-\beta)K_T} \\
	\label{constraint_6}
	P_{k,i}\geq a_{k,i}\sigma^2 \Gamma ~~\substack{\forall i \in [0, N_D], \\
	\forall k \in [1, K_T]}
\end{IEEEeqnarray}

The objective function in (\ref{obj_function}) minimizes the number of deployed ABSs and positions them in selected locations to enhance performance through maximizing the received power levels at the users, which maps directly to increasing the total network bit rate. 
A configurable parameter $\lambda$ is introduced to control the balance between maximizing the network performance quality and minimizing the number of deployed ABSs. For example, when the value of $\lambda$ is small, the number of required ABSs would be high to achieve better network quality. Therefore, in our performance results, we have tested different values of $\lambda$ and selected a relatively large value to obtain desirable results with balance between cost (reflected by number of UAVs) and performance (reflected by network bit rate). 

The first constraint represented in (\ref{constraint_1}) ensures that a user will only be linked to an available ABS. The second constraint in (\ref{constraint_2}) enforces each user to associate with only one site, either TBS or ABS. The third and fourth constraints in (\ref{constraint_3}) and (\ref{constraint_4}), respectively, guarantee that the number of users linked to each site (TBS or ABS) do not exceed the maximum capacity limits. The constraint in (\ref{constraint_5}) requires the total number of covered users to be at least ($1-\beta$) of the total number of users, where $\beta$ is the desired outage probability. Finally, the constraint in (\ref{constraint_6}) ensures a minimum SNR threshold for a user to associate with its serving site. 

This problem resembles the capacitated facility location~(CPL) problem that  is known to be NP-hard~\cite{mihelic2004,farahani2010multiple,silva2007capacitated}. The CPL problem aims to pick $F$ facilities with capacity $X$ while minimizing the sum of distances between the demand points and corresponding facilities. We can map the CPL problem to our problem by setting $K_B$ and $K_D$ to $0$ and $X$, respectively, in addition to setting $\beta$ and $\Gamma$ to $0$ so we consider all demand points and minimize the sum of distances. Therefore, it would be highly complex to solve the formulated optimization problem except for small scale scenarios. To deal with this, we present next a greedy approach and demonstrate its ability to generate close-to-optimal results with much lower computational complexity.  

\subsection{Greedy Approach}

In this section, we present a greedy approach for UAV deployment composed of two phases that can achieve results close to the solution of the optimization problem in~(\ref{obj_function}). The aim of this approach is to develop a close-to-optimal benchmark to evaluate the performance of the low complexity algorithm proposed in Section~\ref{sec:force} in relatively large scale network scenarios. 

The first phase, referred to as \textit{selection phase}, determines the minimum number of needed ABSs. The second phase, referred to as \textit{association phase}, finds the best locations to position the ABSs in a way to maximize the received power level at the users. The solution also accounts for the capacity constraints of the TBS and ABSs. 
The selection phase starts by assigning all uncovered users to the nearest available site whether TBS or ABS and then computes a score for each site $i$ based on the following function:
\begin{equation}\label{eq:score}
 \frac{\displaystyle\sum\limits_{k\in {V_i}} { P_{k,i}}}{|V_i|}-\frac{P_0}{|V_i|},             
\end{equation}

\noindent where $ {V_i}$ is the set of users within the coverage range of site $i$, $|V_i|$ denotes the cardinality of ${V_i}$, and $P_0$ is the power received over a given reference distance. This score function attempts to order sites according to the average received power of users within their respective coverage area. Starting with the site having the highest score, its corresponding $|V_i|$ users are eliminated from the list of uncovered users as long as the site capacity constraint is maintained. The process repeats for the next sites in decreasing order of scores  until the outage probability constraint is met or the list of uncovered users becomes empty. Algorithm ~\ref{algorithm} describes the various steps of this phase. 

\begin{algorithm}
\caption{Greedy Approach: Selection Phase}
\label{algorithm}
\begin{algorithmic}[1]
\State $\bf{Input:}$
\State $D\gets Set \hspace{1 mm} of \hspace{1 mm} all \hspace{1 mm} possible  \hspace{1 mm} sites \hspace{1mm} (ABSs \hspace{1 mm} and \hspace{1 mm} TBS) $
\State $U\gets Set \hspace{1 mm} of \hspace{1 mm} all \hspace{1 mm} users \hspace{1 mm}  $
\State $\bf{Output:}$
\State $M\gets Set \hspace{1 mm} of \hspace{1 mm} selected \hspace{1 mm} sites$
\Procedure{Selection\textendash Phase}{}
 \State $min\_coverage\gets (1-\beta)K_T$
 \State $users\_covered\gets 0$
\State $V_i\gets Set \hspace{1 mm} of \hspace{1 mm} users\hspace{1 mm} within\hspace{1 mm} coverage\hspace{1 mm} of\hspace{1 mm} site\hspace{1 mm} i \gets \Phi$ 
\State $M\gets \Phi$
\While{ ( $users\_covered$ \textless $min\_coverage$ )}
  \For{each $i \in D$}
		\If{  ($i \hspace{1mm} not\hspace{1 mm} used )$ }
		\State $calculate\hspace{1mm} score \hspace{1mm} S_i\hspace{1mm} of\hspace{1mm} i \hspace{1mm} according\hspace{1mm} to \hspace{1mm} \eqref{eq:score}$
        \EndIf
  \EndFor
\State $P\gets site \hspace{1 mm} with \hspace{1 mm} highest \hspace{1 mm} score$
\State $M\gets M \cup P  $
\State $V_P\gets \Phi  $

\For{each $u \in U$}
		\If{  ($|V_P| \geq \hspace{1 mm} capacity \hspace{1 mm} of \hspace{1 mm} P $ ) }	
	    \State break
        \EndIf
	   
	    \If{  ($u$ is within coverage range of $P$ ) }
	      \State $U\gets U-u  $
	      \State $V_P\gets V_P \cup u   $
        \EndIf
\EndFor

\State $users\_covered\gets users\_covered + |V_P| $

\EndWhile 
\break
\Return $M$
\EndProcedure
\end{algorithmic}
\end{algorithm}

\begin{algorithm}
\caption{Greedy Approach: Association Phase}
\label{algorithm2}
\begin{algorithmic}[1]
\State \bf{Input:}
\State $U\gets Set \hspace{1 mm} of \hspace{1 mm} all \hspace{1 mm} users \hspace{1 mm}  $
\State $M\gets Set \hspace{1 mm} of \hspace{1 mm} selected \hspace{1 mm} sites$
\State \bf{Output:}
\State $All\hspace{1 mm} sets \hspace{1 mm}V_m$ \State $V_m\gets Set \hspace{1 mm} of \hspace{1 mm} users\hspace{1 mm} associated\hspace{1 mm} to\hspace{1 mm} site\hspace{1 mm} m, \hspace{1 mm} where \hspace{1 mm} m \in M$
\Procedure{Association\textendash Phase}{}
\For{each $m \in M$}
		\State $V_m\gets \Phi  $
\EndFor
\For{each $u \in U$}
	\State $m\gets closest \hspace{1 mm} ABS, \hspace{1 mm} to\hspace{1 mm} user\hspace{1mm}u \hspace{1mm}(if\hspace{1mm} capacity\hspace{1mm}is \break\indent\hspace{3 mm} not\hspace{1mm} exceeded)$
		 \State $V_m\gets V_m \cup u   $
		 \State $U\gets U-u  $
\EndFor
\EndProcedure
\end{algorithmic}
\end{algorithm}

Upon completing the selection phase, the minimum number of ABSs is determined. The second phase works on associating users to the selected ABSs based on received power level in order to maximize the network bit rate. This continues until the capacity constraint of each site is reached. The association phase is summarized in Algorithm~\ref{algorithm2}. 

\begin{figure}[t!]
	\centering
	\includegraphics[width=0.53\textwidth]{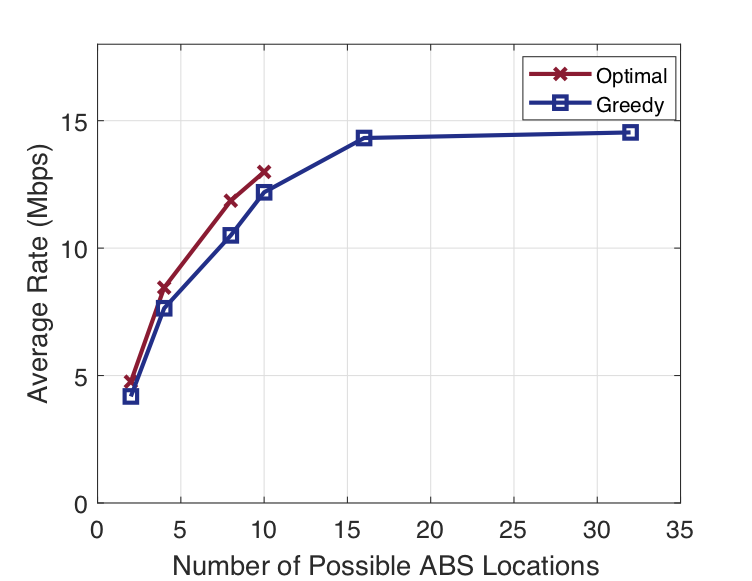}
	\caption{Average bit rate in Mbps for the greedy approach compared to the optimal solution, as a function of the number of possible ABS locations.}
	\label{fig:RateVsABS}
\end{figure}

\begin{figure}[t!]
	\centering
	\includegraphics[width=0.53\textwidth]{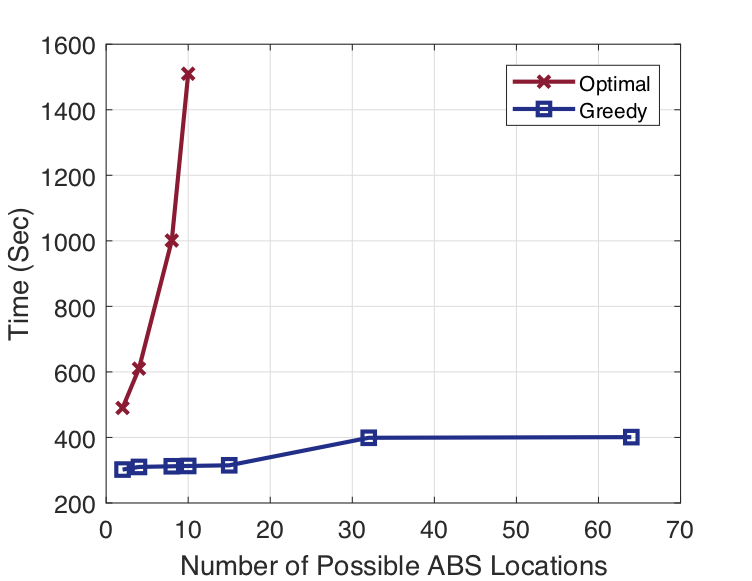}
	\caption {Execution time of the optimal solution versus the  greedy approach as a function of the number of possible ABS locations.}
	\label{fig:TimeVsABS}
\end{figure}

In order to demonstrate the effectiveness of the presented greedy approach, we solve the formulated optimization problem in~(\ref{obj_function}) using Matlab's branch and bound optimization toolbox and compare the optimal results to the greedy solution for several scenarios. We use the same system parameters presented in Section~\ref{sec:results} assuming $K_T=200$~users in $100m\times100m$ area. 
Fig.~\ref{fig:RateVsABS} reflects the improvement in user performance as a function of the number of possible ABS locations $N_D$. This figure demonstrates similar behavior between the optimal solution and the greedy approach. Increasing the number of possible ABS locations, however, leads to larger search space that eventually impacts the execution time to generate the optimal solution as demonstrated in Fig.~\ref{fig:TimeVsABS}.  



\begin{figure}[t!]
	\centering
	\includegraphics[width=0.53\textwidth]{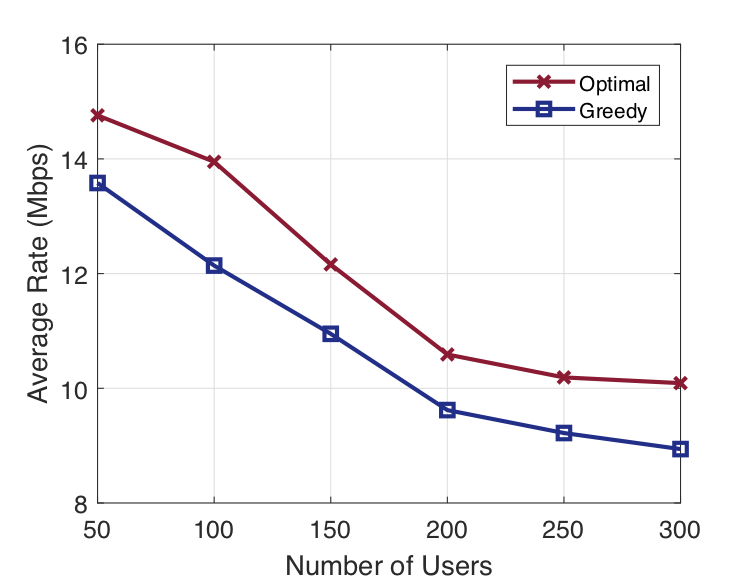}
	\caption{Average bit rate in Mbps for greedy approach compared to the optimal solution, as a function of the number of users assuming six possible ABS locations in $100m\times100m$ area. }
	\label{fig:RateVsUsers}
\end{figure}

The two-phase greedy approach is shown to be an efficient alternative to generate near-optimal results with low complexity for relatively large network scenarios. In Fig.~\ref{fig:RateVsUsers}, the number of possible ABS locations is fixed to six and the number of users is increased within the given geographical area. As the number of users grows, the average user bit rate drops as expected and both solutions, optimal and greedy, exhibit relatively close performance. 

\section{UAV Deployment Algorithm Based on Electro- Static Forces With Continuous Search Space}
\label{sec:force}

The problem formulation in Section~\ref{sec:formulation} assumes a finite set of possible locations on a grid for deploying ABSs. In practice, however, UAVs can hover in continuous 3D space and can stop at any position that provides enhanced network performance. In this section, we propose a novel and effective algorithm for UAV deployment in 3D continuous space based on the notion of electrostatic systems with repulsion and attraction forces. 

Based on the adopted system model, we aim at deploying the least number of ABSs to offload traffic from an existing TBS whenever a sudden increase in traffic demand occurs. We  utilize an unsupervised learning approach based on electrostatic forces to make ABSs park at optimized spots. 

To apply the law of electrostatics, we assign positive dynamic charges to ABSs and fixed negative charges to users. 
Hence, forces among ABSs are repulsive so they tend to get away from each other while they get attracted to users, being carriers of opposite charges. Repulsive and attractive forces are calculated according to Coulomb's Law that quantifies the force between two electrically charged points.
Based on the assigned charges, a repulsive force forms between two ABSs, $i$ and $j$, as follows: 
\begin{equation}
    \vec{F}_{d_i,d_j}=\frac{Q_{d_i}Q_{d_j}}{T^2}\dfrac{\vec{c}_{d_i}-\vec{c}_{d_j}}{||\vec{c}_{d_i}-\vec{c}_{d_j}||},
    \label{drone_force}
\end{equation}

\noindent where $Q_{d_i}$ and $Q_{d_j}$ denote charges of ABS $i$ and ABS $j$, respectively, $\vec{c}_{d_i}$ and $\vec{c}_{d_j}$ denote their coordinate vectors in 3D space, and $T$ denotes the distance between them. As for the attractive force $\vec{F}_{d_i,u}$ that exists between an ABS $i$ and a user $u$, it is similarly calculated using charge $Q_u$ assigned to each user $u$ and its coordinate vector $\vec{c}_u$. 


The total force exerted on each ABS is then evaluated as the summation of forces generated by each of the other ABSs in addition to forces generated by the users. Hence, the force applied on each ABS $d_i$ is calculated as follows: 
\begin{equation}
    \vec{F}_{d_i}=\sum\limits_{d_j \in D, j \neq i}\vec{F}_{d_i,d_j} + \sum\limits_{u \in U}\vec{F}_{d_i, u}.
    \label{drone_all}
\end{equation}

As stated earlier, users are assigned static negative charges while ABSs are assigned dynamic positive charges. We set each user charge $Q_u$ to 1 and each ABS charge to $Q_{d_i}$ which is regularly updated inversely proportional to the associated number of users. To this end, ABSs with larger number of users generate weaker attractive force to new users enabling load balancing among ABSs. Thus, the charge assigned to ABS $i$ is evaluated in terms of the number of users associated with it as follows: 
\begin{equation}
    Q_{d_i}=\frac{\alpha}{k_i+1},
    \label{charge}
\end{equation}
\noindent where $\alpha$ is a value between $0$ and $1$ and $k_i$ is the number of users associated with ABS $i$.

Assigning virtual electric charges causes an electric field to form between ABSs and users, and among ABSs. The ABSs would approach users due to opposite charges and repel from each other due to similar charges. They keep on moving in the space until electrostatic equilibrium is achieved leading to fixed locations in 3D space; equilibrium is achieved when minimal change is observed over multiple iterations. The ABSs are assumed to move in certain step sizes where the direction in which they move is calculated according to the sum of forces exerted on each of them. Using a pre-configured fixed step size, the new coordinate vector of an ABS at time $\tau+1$ is calculated as follows: 
\begin{equation}
    \vec{c}_{d_i}^{\,\tau+1}=\vec{c}_{d_i}^{\,\tau}+\eta\frac{\vec{F}_{d_i}}{||\vec{F}_{d_i}||},
    \label{position}
\end{equation}
\noindent where $\eta$ is the step size and $\tau$ is the time iteration index.

\begin{figure}[t!]
	\centering
	\includegraphics[width=0.53\textwidth]{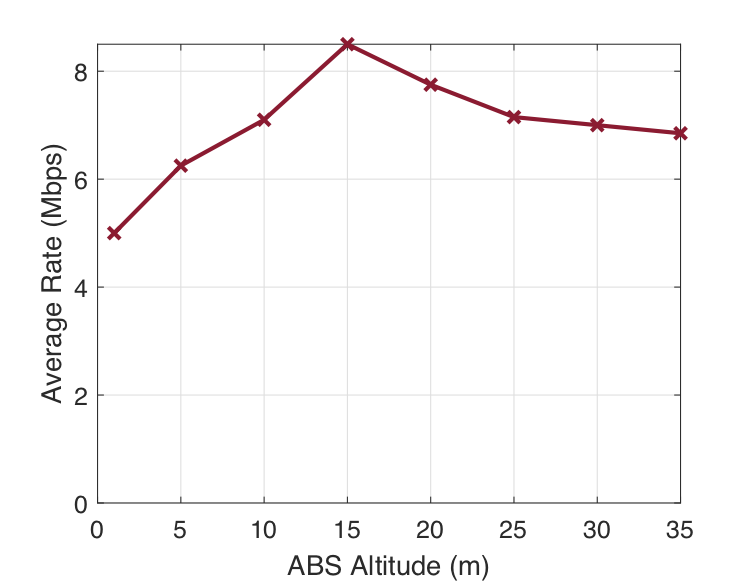}
	\caption{Average bit rate for different ABS heights.}
	\label{fig:Bh}
\end{figure}

\begin{figure}[t!]
\centering
\includegraphics[width=0.53\textwidth]{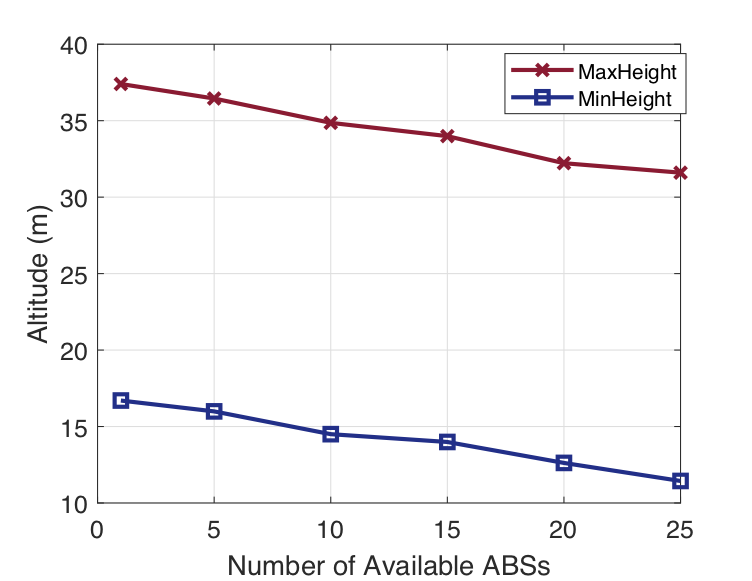}
\caption{ Minimum and maximum ABS heights in an urban environment assuming $100m\times100m$ area. }
\label{fig:hh}
\end{figure}

The altitude of a given ABS impacts network performance due to existing trade offs. For example, Fig.~\ref{fig:Bh} demonstrates how the height at which the ABS is deployed affects the achieved user bit rate. Higher altitude leads to better bit rates due to the increase in coverage area. However, as the altitude exceeds a certain level, the achieved bit rate starts to drop as the received power level weakens notably due to channel path loss. In order to limit the 3D continuous search space, we bound the possible height between minimum and maximum levels within which individual ABSs can be placed. The minimum height is determined by assuming all available ABSs are in operation to cover the given area. As the height at which an ABS is deployed increases, the coverage area increases due to improved line of sight but then starts to drop due to weakened received signal strength. The maximum height is calculated in a way for a single ABS to cover the largest area based on the selected path loss model. In Fig.~\ref{fig:hh}, we vary the number of available ABSs and plot the minimum and maximum heights at which they can operate assuming an urban environment. 

\begin{figure}[t!]
\centering
\includegraphics[width=0.51\textwidth]{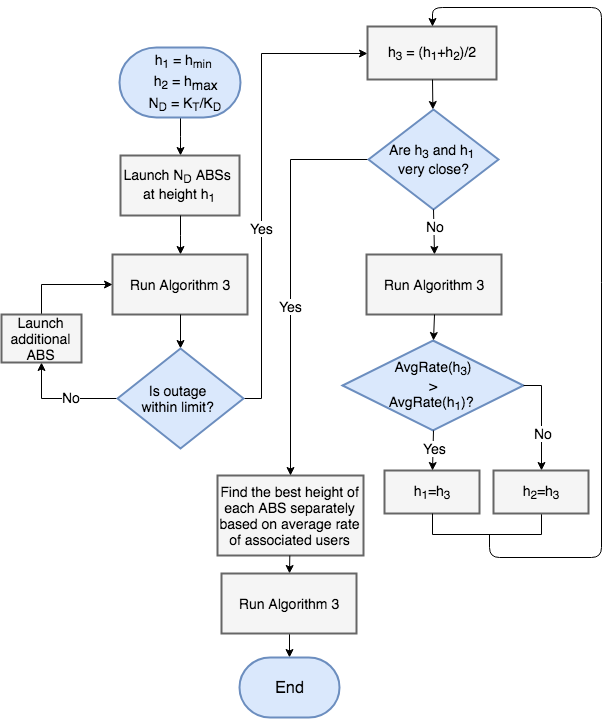}
\caption{Proposed Force3D algorithm flowchart. }
\label{fig:flowchart} \vspace{0.4cm}
\end{figure}

Fig.~\ref{fig:flowchart} and Algorithm~\ref{algorithm2d} describe the complete operation of our proposed 3D positioning algorithm based on electrostatic forces, referred to as Force3D hereinafter. Fig.~\ref{fig:flowchart} depicts the sequence of steps taken by Force3D to find the best position that each ABS parks at. Initially, the minimum number of ABSs is launched randomly on one plane positioned at the minimum height. ABSs and users are assigned their charges and ABSs relocate as per Algorithm~\ref{algorithm2d}, and then the resulting user outage is evaluated. Additional ABSs are launched until the outage probability drops below the set target level. Afterwards, the algorithm moves the plane of ABSs between the minimum and maximum heights until it achieves the best average user rate. In our implementation, we used binary search to search for the best height but other implementations are also possible. Next, we lock the users associated with every ABS and move the ABS vertically to search for the position that provides the best average rate to the associated users. This latter process is repeated for all ABSs independently. At that point, each ABS is positioned at a different height in the 3D space. Finally, Algorithm~\ref{algorithm2d} is executed as a final step to refine the 3D positions based on repulsion and attraction forces depending on the allocated charges. It is worth noting that Algorithm~\ref{algorithm2d} keeps all ABSs on the same plane if they so started. 


\begin{algorithm}[t!]
\caption{Placement Algorithm for Multiple UAVs Based on Electrostatic Forces}
\label{algorithm2d}
\begin{algorithmic}[1]
\State $\bf{Input:}$
 \State $D\gets Set \hspace{1 mm} of \hspace{1 mm} N \hspace{1 mm} ABSs$
 \State $U\gets Set \hspace{1 mm} of \hspace{1 mm} users $
\State $\bf{Output:}$
 \State $C\gets Set \hspace{1 mm} of \hspace{1 mm} coordinate\hspace{1 mm} vectors\hspace{1 mm} of\hspace{1 mm} all\hspace{1 mm} ABSs$
\Procedure{Placement Algorithm}{}
\While{ ( Equilibrium not achieved )}
  \For{each $i \in D$}
  		\State $k_i\gets 0  $ {\hspace{1mm}//initial number of users associated to \break\indent\indent\indent\indent ABS $i$}
  		\State $Q_{d_i}\gets 0  ${\hspace{1mm}//initial charge of ABS $i$}
        \State $\vec{F}_{d_i}\gets \vec{0}  $ {\hspace{1mm}initial force applied on ABS $i$}
		\While{($k_i$ \textless $K_D $ )}
		    \State {associate nearest user $u$ to ABS $i$} 
		    \State $U\gets U-u  $
	        \State $k_i\gets k_i + 1   $
		\EndWhile 
	    \State $Q_{d_i}\gets \frac{\alpha}{k_i+1}   $
	          
  \EndFor
\For{each $i \in D$}
\For{each $v \in D \cup U$}
  		\State $\vec{F}_{d_i}\gets \vec{F}_{d_i} + \vec{F}_{d_i,v} $ 
	  \EndFor        
  \EndFor
  
  \For{each $i \in D $}
	\State $\vec{c}_i\gets \vec{c}_i + \eta \frac{\vec{F}_{d_i}}{||\vec{F}_{d_i}||}  $         
  \EndFor

\EndWhile 
\break
\Return $C$
\EndProcedure
\end{algorithmic}
\end{algorithm}

  


\section{Simulation Results and Performance Analysis}
\label{sec:results}

In this section, we generate and analyze simulation results to study the behavior of the proposed Force3D algorithm as compared to the greedy approach presented in Section~IV.A and to recent related work from the literature. A wide range of simulations are conducted with averaging over large-enough number of iteration assuming the system parameters presented in Table~\ref{tab:constants_used}. 

\begin{table}[t!]
	\caption{System parameters for simulation results}
	\centering
	\begin{tabular}{ |l|l| }
		\hline
		Parameter      & Value \\ 
		\hline
		TBS transmit power $P_B$      & 20 Watts\\ 
		ABS transmit power $P_D$      & 5 Watts\\
		TBS maximum capacity $K_B$      & 50\\
		ABS maximum capacity $K_D$      & 20\\
		path loss constant $\kappa$   & -30 dB\\
		path loss exponent $a $   & 4 \\
		reference distance $d_0$      & 1 m \\
		thermal noise power $\sigma^2$ & $10^{-6}$ Watts \\
		SNR threshold $\Gamma$ &  2 dB \\
		carrier frequency $f_c$ &  2.5 GHz \\
		$\mu$ & 9.61 \\
	    $\gamma$ & 0.16 \\
	    $\eta_{LoS}$ & 1 \\
	    $\eta_{NLoS}$ & 20\\
	    $\eta$ & 0.4 \\
		\hline
	\end{tabular}
	\label{tab:constants_used}
\end{table}

\begin{figure}[t!]
	\centering
	\includegraphics[width=0.53\textwidth]{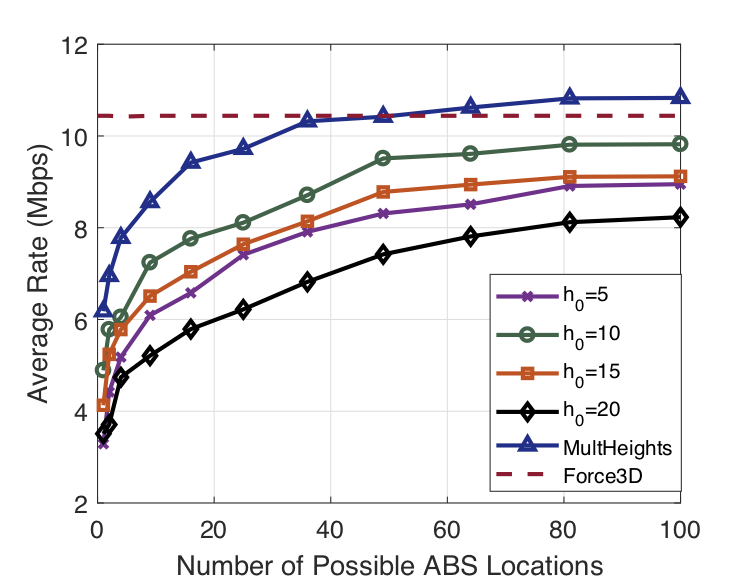}
	\caption{Average user bit rate achieved by the proposed Force3D algorithm compared to the close-to-optimal greedy approach.}
	\label{fig:3d}
\end{figure}

In Fig.~\ref{fig:3d}, the proposed Force3D algorithm with varying UAV heights is compared to the optimized solution generated by the greedy approach with fixed UAV heights (represented by the parameter $h_0$ in meters). It is shown that the ABS height plays a major role in the overall network performance. We also compare with the the greedy approach assuming ABSs are placed at varying heights (denoted as \textit{multiheights}); as expected, this achieves better rates than when all ABSs are positioned at the same height for the same number of possible locations. 

Our proposed Force3D algorithm resulted in higher bit rates than the greedy approach  with fixed heights for all cases, even when the number of possible UAV locations grew to as high as 100 over the given limited geographical area. The \textit{multiheight} case led to slightly better results than Force3D starting from 50 possible UAV locations; this demonstrates that the proposed low complexity Force3D algorithm can achieve results that are close to optimized solutions. 



\begin{figure}[t!]
	\centering
	\includegraphics[width=0.53\textwidth]{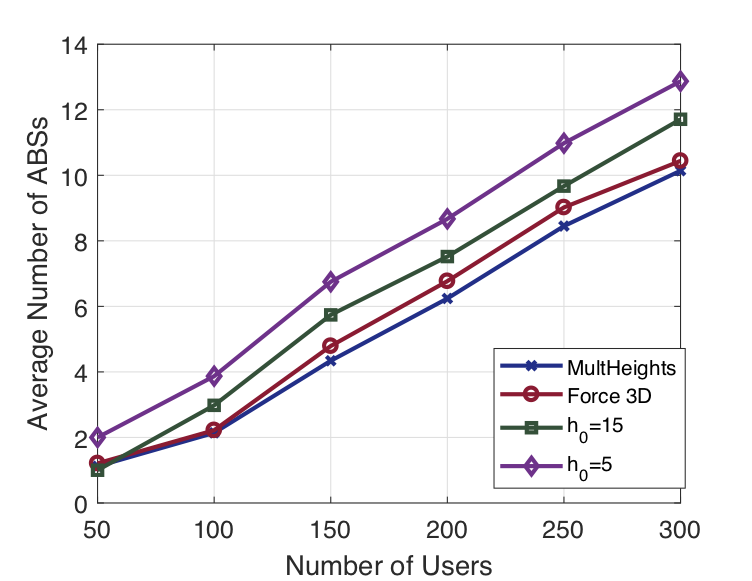}
	\caption{Average number of ABSs required to cover the users in the given geographical area.}
	\label{fig:nmDrone}
\end{figure}

Fig.~\ref{fig:nmDrone} presents the needed number of ABSs with respect to the available number of users. Again, Force3D achieves close results compared to the greedy approach with multiple heights, and better results than the greedy approach when all ABSs are placed at the same height. 

\subsection{Performance Comparison with Existing Work from the Literature}

In this subsection, we compare the performance of our proposed Force3D algorithm with recent work published in~\cite{7762053}  that also aims to minimize the number of UAV-mounted base stations to cover a given area. The approach in~\cite{7762053} uses a polynomial-time algorithm that places UAVs sequentially in a spiral manner starting from the area perimeter towards the center until all ground users are covered. We note here that the approach only works on two dimensional placement of the ABSs and, thus, it will be referred to \textit{Spiral2D}.   

\begin{figure}[t!]
	\centering
	\includegraphics[width=0.52\textwidth]{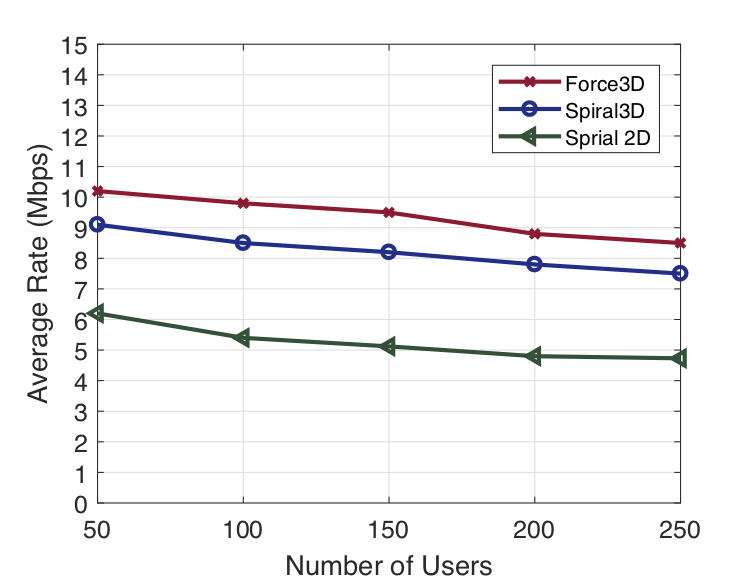}
	\caption{Average user bit rate for Force3D versus Spiral2D and Spiral3D assuming uniform distribution of users over the given area.}
	\label{fig:compareUniform}
\end{figure}

\begin{figure}[t!]
	\centering
	\includegraphics[width=0.53\textwidth]{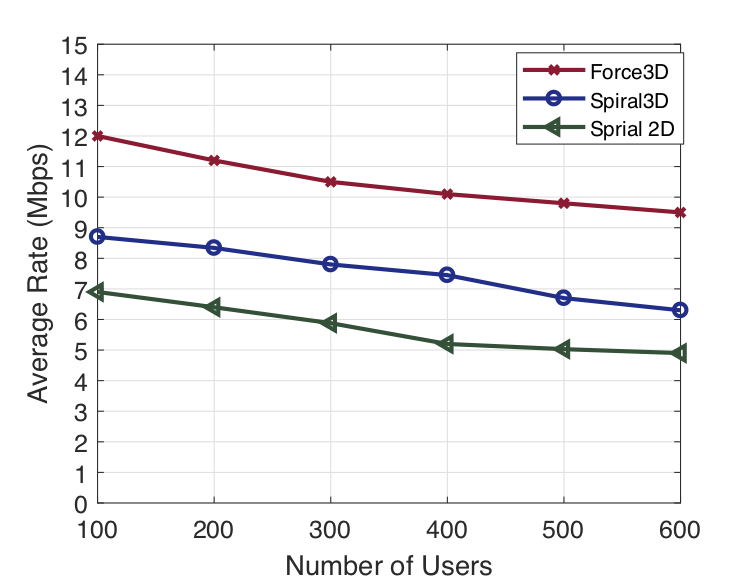}
	\caption{Average user bit rate for Force3D versus Spiral2D and Spiral3D assuming non-uniform distribution of users with dense hot spots over the given area.}
	\label{fig:compareDense}
\end{figure}

\begin{figure}[t!]
	\centering
	\includegraphics[width=0.53\textwidth]{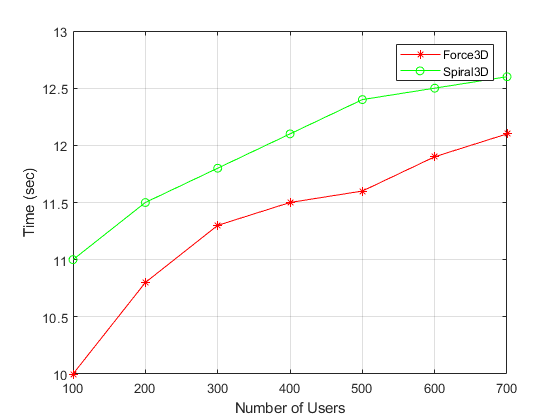}
	\caption{Average execution time comparison for Force3D versus Spiral3D as a function of the number of users.}
	\label{fig:compareComplexity}
\end{figure}

Fig.~\ref{fig:compareUniform} compares the performance of our proposed Force3D approach with Spiral2D for varying number of users uniformly distributed in an area of $100m\times100m$. Force3D consistently achieves around 67\% increase in average bit rate as compared to Spiral2D. To make a fair comparison, we have modified the implementation of Spiral2D to operate in a 3D search space similar to Force3D; yet, Force3D is shown to achieve better bit rates. The performance gap is further amplified in areas with dense hot spots (non-uniform user distribution) as shown in Fig.~\ref{fig:compareDense}, where the percentage bit rate increase is almost 72\% compared to Spiral2D and 33\% compared to Spiral3D.

We have also compared the complexity of Force3D versus Spiral3D by monitoring the execution time in seconds on the same desktop with a given set of input parameters. Results are shown in Fig.~\ref{fig:compareComplexity} as a function of the number of users; they demonstrate that the Force3D algorithm is notably more efficient than Spiral3D in terms of execution time. In summary, the proposed Force3D algorithm achieves favorable performance in terms of both performance and complexity.  




\section{Conclusions}\label{sec:conc}
We addressed the problem of deploying a fleet of UAV-mounted base stations to cover users in a dense network for the purposes of traffic offloading and/or network recovery. We formulated the problem with 3D discrete search space as a mixed integer program and proposed a greedy approach that achieves close-to-optimal results. Then, we developed and evaluated a novel, effective and low complexity algorithm based on the notion of electrostatic forces to position aerial base stations in 3D continuous search space. Extensive simulation results are presented to demonstrate the effectiveness of the proposed algorithm compared to the greedy approach and to recent related work from the literature. The performance results included evaluation of both network bit rates and execution time complexity, for several scenarios with uniform and non-uniform distribution of users. 

\bibliography{paper}

\end{document}